\documentclass{ws-procs975x65}

\begin{document}

\title{Cosmic-Ray Proton to Electron Ratios}

\author{M. PERSIC$^*$}

\address{INAF/Osservatorio Astronomico di Trieste\\
via G.B.Tiepolo 11, I-34143 Trieste, Italy\\
$^*$E-mail: persic@oats.inaf.it\\
www.oats.inaf.it}

\author{Y. REPHAELI}

\address{School of Physics and Astronomy, Tel Aviv University\\
Tel Aviv 69978, Israel\\
E-mail: yoelr@wise.tau.ac.il}

\begin{abstract}
A basic quantity in the characterization of relativistic particles is the proton-to-electron 
(p/e) energy density ratio. We derive a simple approximate expression suitable to estimate 
this quantity, $U_{\rm p}/U_{\rm e} = (m_{\rm p}/m_{\rm e})^{(3-q)/2}$, valid when a nonthermal 
`gas' of these particles is electrically neutral and the particles' power-law spectral indices 
are equal -- e.g., at injection. This relation partners the well-known p/e number density ratio 
at 1\,GeV, $N_{\rm p}/N_{\rm e} = (m_{\rm p}/m_{\rm e})^{(q-1)/2}$.
\end{abstract}


\bodymatter

\section{Introduction}

The proton-to-electron (p/e) number-density and energy-density ratios are very useful relations in 
cosmic-ray (CR) studies. Simple expressions for these ratios are commonly used, but while the standard 
limiting formula for the first ratio has been derived long ago (e.g., Schlickeiser 2002), there seems to 
be no derivation (to our best knowledge) of a similar expression for the p/e energy density ratio. 
Due to the basic interest in the latter ratio (e.g., in the estimation of the proton energy density from 
the more readily measured electron energy density), it is useful to have a convenient expression also for 
this ratio. This can be easily obtained when the total number of energetic protons very closely equals 
that of energetic electrons.

\section{Number density ratio}

We begin with the common assumption that suprathermal protons and electrons with initial kinetic energy 
$T_0 \simeq 10\,$keV are accelerated to relativistic energies, attaining a power-law spectral distribution 
in momentum, $N_{\rm j}(p) = N_{0, {\rm j}} p^{-q_{\rm j}}$ where j=e,p for, respectively, electrons and 
protons; generally, $q_{\rm e} \neq q_{\rm p}$. If charge neutrality is preserved, the number density of 
each particle species is 
\begin{eqnarray}
n_o ~=~ \int_{T_0}^{\infty} N_{\rm e}(T)\, {\rm d}T ~=~ \int_{T_0}^{\infty} N_{\rm p}(T)\, {\rm d}T \,.
\label{eq:el_neutr}
\end{eqnarray}
From $T = \sqrt{m^2c^4+p^2c^2} - mc^2$, it follows that $p=\sqrt{T^2/c^2+2Tm}$, and ${\rm d}p/{\rm d}T =
(T/c^2+m)(T^2/c^2+2Tm)^{-1/2}$. As $N_{\rm j}(T)=N_{\rm j}[p(T)] {\rm d}p/{\rm d}T$, it follows 
\begin{eqnarray}
N_{\rm j}(T) ~=~ {N_{0, {\rm j}} \over c^2} ~(T+m_{\rm j}c^2)~ \biggl({T^2 \over c^2}+2Tm_{\rm j} 
\biggr)^{-(q_{\rm j}+1)/2}\,.
\label{eq:CR_spectr}
\end{eqnarray}
Inserting Eq.(\ref{eq:CR_spectr}) in Eq.(\ref{eq:el_neutr}) we obtain the normalization of each CR species, 
\begin{eqnarray}
N_{0, {\rm j}} ~=~ n_0 ~(q_{\rm j}-1)~ \biggl[ {T_0^2 \over c^2} +2T_0 m_{\rm j} \biggr]^{(q_{\rm j}-1)/ 2}\,.
\label{eq:CRspectr_norm}
\end{eqnarray}
Because of the assumed electrical neutrality of the primary CRs, we get
\begin{eqnarray}
{ N_{0, {\rm p}} \over N_{0, {\rm e}} } ~=~ 
{ q_{\rm p}-1 \over q_{\rm e}-1 } ~ 
{ \bigl[(T_0/c^2) +2T_0 m_{\rm p}\bigr]^{(q_{\rm p}-1)/ 2} 
      \over 
 \bigl[(T_0/c^2) +2T_0 m_{\rm e}\bigr]^{(q_{\rm e}-1)/ 2} } \,;
\label{eq:CRnorm_ratio}
\end{eqnarray}
if $q_{\rm p}=q_{\rm e}=q$, Eq.(\ref{eq:CRnorm_ratio}) reduces to 
\begin{eqnarray}
N_{0, {\rm p}} / N_{0, {\rm e}}  ~=~ (m_{\rm p} / m_{\rm e})^{(q-1)/ 2} \,.
\label{eq:CRnorm_ratio2}
\end{eqnarray}

The p/e number density ratio is 
\begin{eqnarray}
\zeta(T) ~\equiv~ { N_{\rm p}(T)\,{\rm d}T   \over   N_{\rm e}(T)\,{\rm d}T} \,.
\label{eq:zeta1}
\end{eqnarray}
Inserting Eqs.(\ref{eq:CR_spectr}),(\ref{eq:CRnorm_ratio}) in Eq.(\ref{eq:zeta1}), we obtain 
\begin{eqnarray}
\zeta(T) = 
{ (q_{\rm p}-1) \over (q_{\rm e}-1) } \, 
   { [T_0^2+2T_0m_{\rm p}c^2]^{q_{\rm p}-1 \over 2} \over [T_0^2+2T_0m_{\rm e}c^2]^{q_{\rm e}-1 \over 2} } \, 
{ T^{-{(q_{\rm p}+1) \over 2}} (T+m_{\rm p}c^2) (T+2m_{\rm p}c^2)^{-{q_{\rm p}+1 \over 2}}
       \over 
     T^{-{(q_{\rm e}+1) \over 2}} (T+m_{\rm e}c^2) (T+2m_{\rm e}c^2)^{-{q_{\rm e}+1 \over 2}} } \,; 
\label{eq:zeta2}
\end{eqnarray}
setting $q_{\rm p}=q_{\rm e}=q$ (e.g., at CR injection), Eq.(\ref{eq:zeta2}) yields (Schlickeiser 2002) 
\begin{eqnarray}
\zeta(T)   ~=~
	\left\{ \begin{array}{ll} 
        1 ~~~  & \mbox{ ~ ... ~  $T/c^2 \ll m_{\rm e}$}     \\ 
	\propto \left[T / m_{\rm p}c^2\right]^{(q-1)/ 2} & \mbox{ ~ ... ~  $m_{\rm e} \ll T/c^2 \ll m_{\rm p}$}  \\ 
        \left[m_{\rm p}/ m_{\rm e}\right]^{(q-1)/ 2} & \mbox{ ~ ...~  $T \gg m_{\rm p}c^2$ } \,.
\end{array} \right.
\label{eq:zeta3}
\end{eqnarray}

\section{Energy density ratio}

The p/e energy density ratio is 
\begin{eqnarray}
\kappa(T_0; q_{\rm p}, q_{\rm e}) ~\equiv~ 
{\int_{T_0}^{\infty} N_{\rm p}(T)\,T\,{\rm d}T  \over \int_{T_0}^{\infty} N_{\rm e}(T)\,T\,{\rm d}T} \,.
\label{eq:kappa1}
\end{eqnarray}
Inserting Eqs.(\ref{eq:CR_spectr}),(\ref{eq:CRnorm_ratio}) in Eq.(\ref{eq:kappa1}), we obtain 
\begin{eqnarray}
\lefteqn{
\kappa(T_0; q_{\rm p}, q_{\rm e}) ~=~ 
{ (q_{\rm p}-1) \over (q_{\rm e}-1) } ~ 
   { (T_0^2 + 2T_0 m_{\rm p}c^2)^{q_{\rm p}-1 \over 2} \over (T_0^2 + 2T_0 m_{\rm e}c^2)^{q_{\rm e}-1 \over 2} } 
~ \times 
} 
                \nonumber\\
& & {}  \times ~ 
{ \int_{T_0}^{\infty} T^{-{q_{\rm p}-1 \over 2}} 
                        (T+2m_{\rm p}c^2)^{-{q_{\rm p}+1 \over 2}}
			(T+m_{\rm p}c^2) 
			{\rm d}T 
       \over 
    \int_{T_0}^{\infty} T^{-{q_{\rm e}-1 \over 2}} 
                        (T+2m_{\rm e}c^2)^{-{q_{\rm e}+1 \over 2}}
			(T+m_{\rm e}c^2) 
			{\rm d}T } \,. 
\label{eq:kappa2}
\end{eqnarray}
In Table\,1 we report values of $\kappa$ for several $(q_{\rm p}, q_{\rm e})$ pairs of astronomical interest. 

Denoting the integrands on the top and bottom of the r.h.s. of Eq.(\ref{eq:kappa2}), respectively, 
$f_{\rm p}(T)$ and $f_{\rm e}(T)$, we can rewrite 
$\int_{T_0}^\infty f_{\rm p}(T) {\rm d}T = \int_{T_0}^{m_{\rm p}c^2} f_{\rm p}(T) {\rm d}T + 
\int_{m_{\rm p}c^2}^\infty f_p(T) {\rm d}T$ and 
$\int_{T_0}^\infty f_{\rm e}(T) {\rm d}T = \int_{T_0}^{m_{\rm e}c^2} f_{\rm e}(T) {\rm d}T + 
\int_{m_{\rm e}c^2}^\infty f_{\rm e}(T) {\rm d}T$. 
An approximate {\it estimator} of $\kappa$ can be obtained by considering only proton and electron 
energies exceeding the respective particles' rest mass. Then Eq.(\ref{eq:kappa2}) simplifies into 
\begin{eqnarray}
\kappa(q_{\rm p}, q_{\rm e}) ~\simeq ~
{ (q_{\rm p}-1) \over (q_{\rm e}-1) } ~
{ (q_{\rm e}-2) \over (q_{\rm p}-2) } ~ 
{ (2T_0m_{\rm p}c^2)^{q_{\rm p}-1 \over 2} \over (2T_0m_{\rm e}c^2)^{q_{\rm e}-1 \over 2} } ~
{ (m_{\rm p}c^2)^{2-q_{\rm p}} \over (m_{\rm e}c^2)^{2-q_{\rm e}} } \,;& {}
\label{eq:kappa3}
\end{eqnarray}
if $q_{\rm p}=q_{\rm e}=q$ (e.g., at CR injection), Eq.(\ref{eq:kappa3}) reduces to 
\begin{eqnarray}
\kappa(q) ~ \simeq ~ 
\biggl( {m_{\rm p}  \over m_{\rm e}} \biggr)^{(3-q)/2} \,.
\label{eq:kappa4}
\end{eqnarray}
%

\begin{table}
\tbl{Proton-to-electron energy density ratios$^{[a]}$.}
{\begin{tabular}{@{}ccccc@{}}
\hline
\noalign{\smallskip}

$\overline{~q_{\rm p} ~~~~q_{\rm e} \,~~~~~\kappa~~}$ & ~~
$\overline{~q_{\rm p} ~~~~q_{\rm e} \,~~~~~\kappa~~}$ & ~~
$\overline{~q_{\rm p} ~~~~q_{\rm e} \,~~~~~\kappa~~}$ & ~~
$\overline{~q_{\rm p} ~~~~q_{\rm e} \,~~~~~\kappa~~}$ & ~~
$\overline{~q_{\rm p} ~~~~q_{\rm e} \,~~~~~\kappa~~}$ \\

\noalign{\smallskip}
\hline
\noalign{\smallskip}


2.0 ~~2.0 ~~25.8 &~~ 2.1 ~~2.0 ~~9.84 &~~ 2.2 ~~2.0 ~~4.18 &~~ 2.3 ~~2.0 ~~1.97 &~~ 2.4 ~~2.0 ~~1.01 \\
2.0 ~~2.1 ~~62.8 &~~ 2.1 ~~2.1 ~~23.9 &~~ 2.2 ~~2.1 ~~10.2 &~~ 2.3 ~~2.1 ~~4.79 &~~ 2.4 ~~2.1 ~~2.46 \\
2.0 ~~2.2 ~~119\,&~~ 2.1 ~~2.2 ~~45.3 &~~ 2.2 ~~2.2 ~~19.3 &~~ 2.3 ~~2.2 ~~9.06 &~~ 2.4 ~~2.2 ~~4.66 \\
2.0 ~~2.3 ~~189\,&~~ 2.1 ~~2.3 ~~72.0 &~~ 2.2 ~~2.3 ~~30.6 &~~ 2.3 ~~2.3 ~~14.4 &~~ 2.4 ~~2.3 ~~7.40 \\
2.0 ~~2.4 ~~269\,&~~ 2.1 ~~2.4 ~~102\,&~~ 2.2 ~~2.4 ~~43.6 &~~ 2.3 ~~2.4 ~~20.5 &~~ 2.4 ~~2.4 ~~10.5 \\
2.0 ~~2.5 ~~357\,&~~ 2.1 ~~2.5 ~~136\,&~~ 2.2 ~~2.5 ~~57.8 &~~ 2.3 ~~2.5 ~~27.2 &~~ 2.4 ~~2.5 ~~14.0 \\
2.0 ~~2.6 ~~451\,&~~ 2.1 ~~2.6 ~~172\,&~~ 2.2 ~~2.6 ~~73.1 &~~ 2.3 ~~2.6 ~~34.4 &~~ 2.4 ~~2.6 ~~17.7 \\
2.0 ~~2.7 ~~551\,&~~ 2.1 ~~2.7 ~~210\,&~~ 2.2 ~~2.7 ~~89.2 &~~ 2.3 ~~2.7 ~~42.0 &~~ 2.4 ~~2.7 ~~21.6 \\
2.0 ~~2.8 ~~654\,&~~ 2.1 ~~2.8 ~~249\,&~~ 2.2 ~~2.8 ~~106\,&~~ 2.3 ~~2.8 ~~49.9 &~~ 2.4 ~~2.8 ~~25.6 \\
2.0 ~~2.9 ~~760\,&~~ 2.1 ~~2.9 ~~289\,&~~ 2.2 ~~2.9 ~~123\,&~~ 2.3 ~~2.9 ~~57.9 &~~ 2.4 ~~2.9 ~~29.8 \\
2.0 ~~3.0 ~~867\,&~~ 2.1 ~~3.0 ~~330\,&~~ 2.2 ~~3.0 ~~140\,&~~ 2.3 ~~3.0 ~~66.1 &~~ 2.4 ~~3.0 ~~34.0 \\

\noalign{\smallskip}
\hline
\end{tabular}}

\noindent
$^{[a]}$ Limits of integration are 10\,keV--100\,TeV.
\smallskip
\end{table}

\section{Discussion} 

The assumption of electric charge neutrality when only the two main relativistic particle species are 
considered is essentially an approximation whose validity rests on the relatively lower abundances of 
other ionic species in comparison with that of protons. 
The energy density of relativistic electrons is readily deduced from spectral radio (synchrotron) 
measurements, whereas there are only few sources for which the proton energy density can be directly 
deduced from $\gamma$-ray measurements of the radiative decay of neutral pions produced in p-p 
collisions. We derived a simple expression to estimate the p/e energy density ratio as a function of 
$q_{\rm p}$ and $q_{\rm e}$, suitable to estimate $U_p$ from the observationally-deduced value of $U_e$ 
-- and to provide a link between protons and electrons at injection (e.g., $q_{\rm p}=q_{\rm e}$).



\begin{thebibliography}{9}
\bibitem{jarl88} R. Schlickeiser, in {\it Cosmic Ray Astrophysics} (Springer-Verlag, 
                 Berlin, 2002), p.\,472.
\end{thebibliography}
\end{document}